\documentstyle[11pt,epsfig]{article}

\newcommand{\pslash}{p \hspace{-0.17cm} / \,}
\newcommand{\kislash}{k \hspace{-0.20cm} / \,}
\newcommand{\kfslash}{k^{\prime} \hspace{-0.30cm} / \;}
\newcommand{\lslash}{l \hspace{-0.18cm} / \,}
\newcommand{\sslash}{S \hspace{-0.23cm} / \;}

\begin{document}

\title{\bf Transverse single spin asymmetries in inclusive \\ 
deep-inelastic scattering}

\author{A.~Metz, M.~Schlegel, and K. Goeke
 \\[0.3cm]
{\it Institut f\"ur Theoretische Physik II,} \\
{\it Ruhr-Universit\"at Bochum, D-44780 Bochum, Germany}}

\date{\today}
\maketitle

\begin{abstract}
\noindent
In inclusive deep-inelastic lepton-hadron scattering multi-photon 
exchange between the leptonic and the hadronic part of the process 
causes single spin asymmetries.
The asymmetries exist for a polarized target as well as a polarized 
incoming or outgoing lepton, if the polarization vector has a component 
transverse with respect to the reaction plane.
The spin dependent parts of the single polarized cross sections are suppressed like $\alpha_{em} \, m_{pol}/Q$ \--- where $m_{pol}$ denotes the mass of the polarized particle \--- compared 
to the leading terms of the cross section for unpolarized or 
double-polarized deep-inelastic scattering.
Both the target and the beam spin asymmetry are evaluated in the 
parton model.
In the calculation only quark-quark correlators are included.
While this approximation turns out to be justified for the lepton spin 
asymmetries, it is not sufficient for the target asymmetry.
\end{abstract}

\noindent
During the last decades an enormous amount of information on the partonic
structure of the nucleon has been extracted from inclusive deep-inelastic 
lepton nucleon scattering (DIS, $l(k) + N(P) \to l(k') + X(P_X)$).
It is well-known that the cross section for this process is fully 
described by four independent structure functions, provided that one
only considers the electromagnetic interaction. 
For instance the unpolarized cross section of inclusive DIS is given by 
\begin{equation} \label{e:sigmaunp}
k^{\prime 0} \, \frac{d\sigma_{unp}}{d^3\vec{k}'} = 
\frac{4 \, \alpha_{em}^2}{Q^4} 
\bigg( x \, y F_{1}(x,Q^2) + \frac{1-y}{y} F_2(x,Q^2) \bigg) \,,
\end{equation}
and contains the two structure functions $F_1$ and $F_2$.
In Eq.~(\ref{e:sigmaunp}) we make use of the standard DIS variables
\begin{equation} \label{e:var}
Q^2 = 2 \, k \cdot k'\,, \qquad
x = \frac{Q^2}{2 \, P \cdot (k - k')} \,, \qquad
y = \frac{P \cdot (k - k')}{P \cdot k} \,.
\end{equation}
Neglecting the nucleon mass $M$ the variables in~(\ref{e:var})
are related by means of $y = Q^2/(xs)$ with $s = 2 P \cdot k$ 
being the squared {\it cm}-energy of the reaction.
Two additional structure functions, often denoted by $g_1$ and $g_2$, 
appear in double-polarized DIS (longitudinal lepton polarization,
and longitudinal or transverse target polarization) 
(see, e.g., Refs.~\cite{jaffe_89,anselmino_95}).
\\
As long as the typically used one-photon exchange approximation is
considered any single spin asymmetry (SSA) is strictly forbidden in 
inclusive DIS due to parity and time reversal 
invariance~\cite{christ_65}.
However, this is no longer true if multi-photon exchange is taken 
into account.
In fact, in inclusive DIS (transverse) SSAs exist if one goes beyond 
the one-photon exchange approximation~\cite{christ_65}. 
The SSAs arise from a specific correlation between a polarization 
vector $S$ of a particle as well as the 4-momenta of the nucleon and 
of the leptons,
\begin{equation} \label{e:corr}
\varepsilon_{\mu\nu\rho\sigma} \, S^{\mu} P^{\nu} k^{\rho} k^{\prime\sigma} \,,
\end{equation}
where $\varepsilon^{\mu\nu\rho\sigma}$ is the totally antisymmetric Levi-Civita 
tensor.
One can readily convince oneself that only those components of $S$ 
contribute to the correlation~(\ref{e:corr}) which are transverse with 
respect to the reaction plane. 
(In the target rest frame, e.g., the reaction plane is given by the 
leptonic plane.)
The vector $S$ can represent the polarization of the nucleon but also
the polarization of the incoming or outgoing lepton.
Since the SSAs turns out to be proportional to the mass of the polarized
particle the target SSA should be the most attractive candidate from the
experimental point of view.
The expression in~(\ref{e:corr}) is a so-called artificial time-reversal odd
correlation.
Artificial time-reversal and ordinary time-reversal differ in the sense 
that in the former case the initial and final state of a reaction are 
not interchanged.
(For a recent discussion on this issue we refer the reader 
to~\cite{sivers_06}.)
In order to generate a correlation of the type~(\ref{e:corr}) a non-zero 
phase (imaginary part) on the level of the amplitude of the process is
required. 
Such a phase can be provided by multi-photon exchange between the
leptonic and the hadronic part of the reaction.
Therefore, there is no argument which forbids the existence of the 
correlation~(\ref{e:corr}) in inclusive DIS.
On the other hand, the SSAs are proportional to the electromagnetic
fine structure constant $\alpha_{em} \approx 1/137$ which may lead to 
relatively small effects.
Indeed, the results of early measurements of the transverse target SSA
at the Cambridge Electron Accelerator~\cite{chen_68} and at the 
Stanford Linear Accelerator~\cite{rock_70} were compatible with zero within 
the error bars.
However, present experiments with their higher precision should be able 
to observe such effects.
\\
We also note that a lot of work has been devoted to transverse SSAs in 
processes like one-hadron inclusive production in hadron-hadron collisions, 
semi-inclusive DIS, and the Drell-Yan process (see, e.g., 
Refs.~\cite{bunce_76,adams_91b,adams_91c,adams_03,adler_05,HERMES_04,COMPASS_05,efremov_84,qiu_91a,sivers_89,collins_92,brodsky_02a,collins_02}).
In particular over the past 4-5 years this field of research has been 
considerably growing.
On the other hand, for decades no measurement/analysis of a transverse SSA 
in inclusive DIS has been performed.
\\
In addition we mention that in elastic lepton scattering off the nucleon 
transverse SSAs were already discussed long ago~\cite{derujula_71}.
(Note also Ref.~\cite{barut_60} where the transverse SSA for elastic 
scattering of two point-like spin-$\frac{1}{2}$ particles was computed.)
On the theoretical side, renewed interest for this observable emerged
recently~\cite{afanasev_02,gorchtein_04,diaconescu_04,pasquini_04,afanasev_04a,afanasev_04b,borisyuk_05a,borisyuk_05b,gorchtein_05a,gorchtein_05b}
because measurements became feasible and non-zero results were 
observed~\cite{wells_00,maas_04}.
Since the elastic scattering is the limit of inclusive DIS for $x \to 1$, 
one certainly can also expect non-vanishing asymmetries in inclusive DIS 
for arbitrary values of $x$.
\\
In this note we compute the transverse SSAs for a polarized incoming
lepton and for a polarized nucleon target in inclusive DIS by considering 
two-photon exchange between the leptonic and the hadronic part 
of the reaction.
The calculation is performed in the framework of the parton model.
(An early phenomenological calculation of the transverse target SSA 
only considered the excitation of the nucleon to the 
$\Delta(1232)$-resonance~\cite{cahn_70}.)
Here we merely take quark-quark correlation functions into account.
In this approach we obtain a well-behaved result for the beam spin 
asymmetry, whereas the target SSA turns out to be infrared (IR)
divergent.
Possibly this divergence can be removed by including in addition
quark-gluon-quark correlators.
The solution of this point requires further work.
\\
It is worthwhile to mention that two-photon exchange may also be at the 
origin of the observed large discrepancy between the outcome of two 
extraction methods --- Rosenbluth separation and polarization transfer --- 
for the electric form factor of the 
proton~\cite{jones_99,guichon_03,blunden_03,rekalo_03}.
Moreover, our work here is related to 
Refs.~\cite{chen_04,gorchtein_04,afanasev_05} in which the two-photon 
exchange contribution to elastic electron scattering off the nucleon was
treated in the parton model.
\\

\noindent
We start by recalling some elements of the collinear parton model.
This approach essentially relies on two ingredients/approximations:
\begin{enumerate}
\item A fast moving hadron looks like a bunch of partons moving in the same
 direction.
 If, e.g., the nucleon has a large light-cone plus-momentum 
 $P^+ = (P^0 + P^3)/\sqrt{2}$ a parton inside the nucleon has a large
 plus-momentum $p^+$ as well.
 The light-cone minus-momentum $p^-$ and the transverse momentum $\vec{p}_T$
 of a given parton are small compared to $p^+$ and are ignored.
\item In the case of a hard process, like inclusive DIS off the nucleon at
 large $Q^2$, the reaction is computed in the impulse approximation, i.e.,
 one considers the reaction rate for the corresponding process with free 
 partons and sums incoherently over the contributions from the different 
 partons.  
 For DIS this means in particular that the virtual photon interacts with a 
 single free quark, while the remaining partons inside the nucleon merely 
 act as spectators of the reaction.
\end{enumerate}
On the basis of the collinear parton model the structure functions 
in Eq.~(\ref{e:sigmaunp}) are given by
\begin{equation} \label{e:unppm}
F_2(x) = 2 x \, F_1(x) = \sum_q e_q^2 \, x f_1^q(x) \,,
\end{equation}
where $f_1^q$ represents the ordinary unpolarized distribution of a quark 
with flavor $q$ in a nucleon.
The summation in~(\ref{e:unppm}) is running both over quarks and antiquarks,
and $e_q$ denotes the quark charge in units of the elementary charge.
The plus-momentum of the quark, which is struck by the virtual photon, is 
specified by means of the relation $p^+ = x P^+$.
The field-theoretical definition of the quark distribution 
reads (see, e.g., Ref.~\cite{collins_81c})
\begin{equation} \label{e:f1}
f_1(x) = \int \frac{d\xi^-}{4\pi} \, e^{i p \cdot \xi} \,
\langle P,S | \bar{\psi}(0) \, \gamma^+ \, \psi(\xi) | P,S \rangle 
 \Big|_{\xi^+ = \xi_T = 0} \,.
\end{equation}
Here we have used the light-cone gauge in which the Wilson-line, connecting 
the two quark fields in~(\ref{e:f1}) and ensuring color gauge invariance of
the operator, disappears.
We also mention that the scale dependence of parton distributions is 
neglected throughout this work since it is irrelevant for the main point 
of the discussion.
\\

\begin{figure}[t]
\begin{center}
\includegraphics[width=10cm]{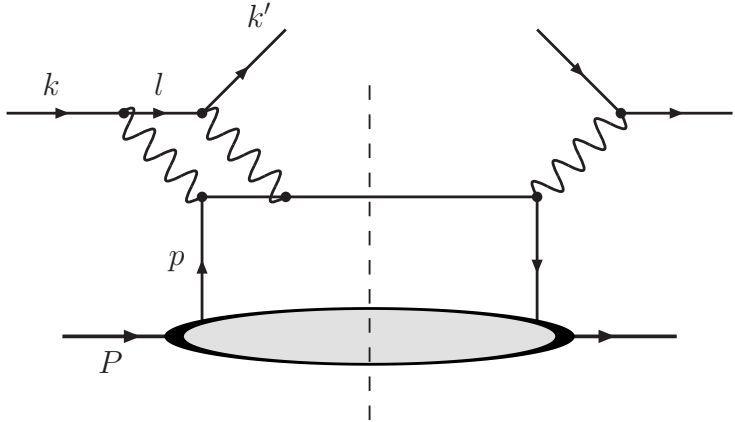}
\end{center}
\caption{Two-photon exchange contribution to inclusive DIS in the parton
model. The Hermitian conjugate diagram, not shown in the figure,
has to be considered as well. A diagram where the ordering of the lower
vertices of the two photons is interchanged (crossed box graph) does not
contribute to the transverse SSA.}
\label{f:twogamma}
\end{figure}

\noindent
Now we turn our attention to the calculation of the transverse SSAs in the 
parton model by discussing in a first step the case of a polarized incoming
lepton. 
To this end we consider the two-photon exchange diagram in 
Fig.~\ref{f:twogamma} together with its Hermitian conjugate.
The so-called crossed box graph, where the lower vertices of the two photons
on the {\it lhs} of the cut in Fig.~\ref{f:twogamma} are interchanged, does 
not contribute to the SSA since it cannot provide an imaginary part.
The second ingredient of the parton model implies that only such diagrams 
are taken into account in which both photons couple to the same quark.  
\\
The diagram in Fig.~\ref{f:twogamma} provides the following contribution
to the squared matrix element of the process,
\begin{eqnarray} \label{e:matrixel}
&& \frac{e^6}{Q^2} \, \frac{1}{i} \int \frac{d^4 l}{(2\pi)^4} \,
 \frac{1}{[(l - k)^2 - \lambda^2 + i \varepsilon] 
          [(l - k')^2 - \lambda^2 + i \varepsilon]  
          [l^2 + i \varepsilon]} \,
 L_{\mu\nu\rho} \; 4\pi W^{\mu\nu\rho} \,,
\\
&& \textrm{with}
\vphantom{\frac{1}{1}}
\nonumber \\
&& L^{\mu\nu\rho} = \frac{1}{2} \, 
 \textrm{Tr} \Big( (\kislash + m) \gamma_5 \sslash \gamma^{\mu} 
                   (\kfslash + m) \gamma^{\nu} (\lslash + m) \gamma^{\rho}
             \Big) \,,
\nonumber \\
&& 4\pi W^{\mu\nu\rho} = \sum_q e_q^3 \pi \, \frac{1}{Q^2} \, 
 \frac{1}{(p + k -l)^2 + i \varepsilon} \, f_1^q(x) \, 
 \textrm{Tr} \Big( \pslash \gamma^{\mu} (\pslash + \kislash - \kfslash)
                   \gamma^{\nu} (\pslash + \kislash - \lslash) \gamma^{\rho} \Big) \,.
\hphantom{aaaa}
\nonumber
\end{eqnarray}
Note that in Eq.~(\ref{e:matrixel}) only the term showing up for a transversely 
polarized incoming lepton is listed.
In order to obtain a non-zero asymmetry one has to work with a finite lepton 
mass $m$. 
When performing the calculation we ignore a term proportional to $m^3$ in
the lepton tensor $L^{\mu\nu\rho}$ and also the mass in the denominator of
the lepton propagator in the loop. 
Both effects are suppressed for large $Q^2$. 
The quark is treated as massless particle.
On the other hand, to avoid a potential IR divergence, a mass $\lambda$ 
is assigned to the photon.
\\
It turns out that in the collinear parton model only the imaginary part 
of the loop-integral in~(\ref{e:matrixel}) survives as soon as one 
adds the contribution coming from the Hermitian conjugate diagram.
This imaginary part can be conveniently evaluated by means of the 
Cutkosky rules.
Here we avoid giving details of the calculation and just quote our 
final result for the spin dependent part of the single polarized cross section,
\begin{equation} \label{e:pollep}
k^{\prime 0} \, \frac{d\sigma_{L,pol}}{d^3\vec{k}'} =  
\frac{4 \, \alpha_{em}^3}{Q^8} \, m \, x \, y^2 \,
\varepsilon_{\mu\nu\rho\sigma} \, S^{\mu} P^{\nu} k^{\rho} k^{\prime\sigma} 
\, \sum_q e_q^3 \, x f_1^q(x) \,.
\nonumber
\end{equation}
At this point several comments are in order.
The result in Eq.~(\ref{e:pollep}) is the leading term in the Bjorken 
limit ($Q^2 \to \infty$, $x$ fixed).
Corrections to this formula are suppressed at least by a factor 
$M/Q$.  
The sign of the spin dependent part of the polarized cross section depends on the charge of the 
lepton which enters to the third power.
The result in~(\ref{e:pollep}) holds for a negatively charged lepton.
(It is interesting to note that in one of the early measurements of the 
target SSA~\cite{rock_70} there is evidence for the expected sign 
change when switching from an electron to a positron beam.)
We have taken the convention $\varepsilon^{0123}=1$ for the Levi-Civita 
tensor.
The spin dependent part of the single polarized cross section behaves like $\alpha_{em} \, m/Q$ relative to the 
unpolarized cross section given in Eq.~(\ref{e:sigmaunp}) (and relative
to the dominant term of the double polarized DIS cross section).
In this context note that the correlation~(\ref{e:corr}) showing up 
in Eq.~(\ref{e:pollep}) is given by
\begin{equation}
\varepsilon_{\mu\nu\rho\sigma} \, S^{\mu} P^{\nu} k^{\rho} k^{\prime\sigma} 
\propto \frac{Q^3}{x \, y} \, \sqrt{1-y}
\end{equation}
in the Bjorken limit.
\\
We emphasize that the expression in Eq.~(\ref{e:pollep}) is IR finite.
Terms proportional to $\ln(Q^2/\lambda^2)$ appearing at intermediate
steps of the calculation cancel in the final result.
In related studies of transverse SSAs in semi-inclusive processes 
a comparable cancellation of IR divergent terms has been observed 
(see, e.g., Refs.~\cite{brodsky_02a,metz_02}).
Because of its IR finiteness the parton model result~(\ref{e:pollep}) 
probably constitutes a reliable estimate of the leading term 
(in the Bjorken limit) of the lepton beam SSA.
Nevertheless, a word of caution has to be added.
At present we have no rigorous proof that other diagrams, not 
included in the parton model approximation, cannot provide a leading 
(and separately IR finite) contribution to the transverse beam SSA.
In fact, one might in particular question the second ingredient of the 
parton model according to which both photons only couple to the 
same quark.
When performing the integration upon the loop-momentum $l$ also photons 
with an arbitrary long wavelength contribute.
Photons with a long wavelength, however, interact with the entire nucleon 
rather than just a single parton.
On the other hand, it is possible that such effects caused by soft 
photon emission in general cancel when computing the lepton SSA.
\\
In connection with the second lepton SSA (polarized lepton in the 
final state) it is sufficient to mention that also this observable is 
IR finite in the collinear parton model.
The calculation is basically a copy of the one for the beam SSA. 
Because the asymmetries are proportional to the mass of the lepton 
they become quite small for electron scattering.
Corresponding measurements of the transverse lepton beam SSA in elastic 
electron nucleon scattering show effects of 
${\cal O}(10^{-6} - 10^{-5})$~\cite{wells_00,maas_04}.
However, in comparison much larger asymmetries can be expected for
polarized muon scattering off the nucleon.
\\

\noindent 
Now we proceed in order to discuss in a second step the transverse target 
SSA.
As we will see below, from a theoretical point of view this observable is 
more challenging than the lepton spin asymmetries.
The main reason for this difference is the twist-3 nature of the target 
asymmetry, whereas the lepton asymmetries, though suppressed like $1/Q$, 
are given by the twist-2 parton density $f_1$. 
\\
We start again by using the collinear parton model.
Also for the target SSA the diagram in Fig.~\ref{f:twogamma} together with 
its Hermitian conjugate is considered.
The calculation proceeds along the lines of the lepton asymmetry, but
here we entirely neglect the lepton mass.
The result for the spin dependent part of the single polarized cross section is now given by
\begin{eqnarray} \label{e:polnuc}
k^{\prime 0} \, \frac{d\sigma_{N,pol}}{d^3\vec{k}'} & = & 
\frac{4 \, \alpha_{em}^3}{Q^8} \, \frac{M \, x^2 \, y}{1-y} \,  
\varepsilon_{\mu\nu\rho\sigma} \, S^{\mu} P^{\nu} k^{\rho} k^{\prime\sigma} 
\\
& & \mbox{} \times 
\bigg( (1-y)^2 \ln \frac{Q^2}{\lambda^2} + y (2-y) \ln y + y (1-y) \bigg) \,
\sum_q e_q^3 \, x g_T^q(x) \,.
\nonumber
\end{eqnarray}
Like in the case of the lepton SSA we have just kept the leading term in the 
Bjorken limit.
As already mentioned the target SSA is a twist-3 effect which is reflected by 
the presence of the twist-3 quark distribution $g_T$ defined through
\begin{equation}
S^i \, g_T(x) = \frac{P^+}{M} \int \frac{d\xi^-}{4\pi} \, e^{i p \cdot \xi} \,
\langle P,S | \bar{\psi}(0) \, \gamma^i \gamma_5 \, \psi(\xi) | P,S \rangle 
 \Big|_{\xi^+ = \xi_T = 0} \,,
\end{equation}
with $i$ denoting a transverse index. 
If one would keep a quark mass then also a term proportional to the 
transversity distribution of the quark would appear in~(\ref{e:polnuc}).
\\
The crucial difference between the result in~(\ref{e:polnuc}) and the  
lepton asymmetry in~(\ref{e:pollep}) is the uncancelled IR divergence
as the photon mass $\lambda \to 0$.
One has to conclude that the collinear parton model is not suitable for 
describing this observable.
As mentioned above it is possible that also diagrams where both photons 
couple to different quarks in the nucleon have to be taken into account 
in order to arrive at an IR finite result.
In addition, it is known for a long time that even in the one-photon exchange 
approximation the collinear parton model is not a sufficient framework to 
describe twist-3 effects in DIS.
Two corrections to the collinear parton model have to be included 
(see, e.g., Ref.~\cite{ellis_82}):
first, the transverse momentum of the struck quark cannot be neglected;
second, also quark-gluon-quark correlators have to be taken into account.
In this note we limit ourselves to the first correction, and leave a detailed
investigation of effects due to quark-gluon-quark correlations inside the
nucleon for future work.
\\
If one considers $p_T$-dependent terms in the quark-quark correlator,
which appear in combination with the imaginary part of the electron-quark
box diagram in Fig.~\ref{f:twogamma}, one arrives at the following result 
for the single polarized cross section,
\begin{eqnarray} \label{e:polnucpt}
k^{\prime 0} \, \frac{d\sigma_{N,pol}}{d^3\vec{k}'} & = & 
\frac{4 \, \alpha_{em}^3}{Q^8} \, \frac{M \, x^2 \, y}{1-y} \,  
\varepsilon_{\mu\nu\rho\sigma} \, S^{\mu} P^{\nu} k^{\rho} k^{\prime\sigma} 
\\
& & \mbox{} \times \int d^2\vec{p}_T \, H(\vec{p}_T^{\;2}) \,
\sum_q e_q^3 \, \Big( x g_T^q(x,\vec{p}_T^{\;2}) 
                   - \frac{\vec{p}_T^{\;2}}{2M^2} \, g_{1T}^q(x,\vec{p}_T^{\;2}) \Big)\,.
\nonumber
\end{eqnarray}
In comparison to the result~(\ref{e:polnuc}) the main new ingredient 
in~(\ref{e:polnucpt}) is the unintegrated ($p_T$-dependent) parton density
$g_{1T}$ which is given by (see, e.g., Ref.~\cite{mulders_95}),
\begin{equation}
\frac{\vec{p}_T \cdot \vec{S}_T}{M} \, g_{1T}(x,\vec{p}_T^{\;2}) = 
\int \frac{d\xi^- \, d^2\vec{\xi}_T}{2(2\pi)^3} \, 
e^{i (p^+ \xi^- - \vec{p}_T \cdot \vec{S}_T)} \,
\langle P,S_T | \bar{\psi}(0) \, \gamma^+ \gamma_5 \, \psi(\xi) | P,S_T \rangle 
 \Big|_{\xi^+ = 0} \,.
\end{equation}
We avoid here discussing the subtle issues of a proper gauge invariant 
definition of $p_T$-dependent parton densities and just refer to the 
literature~\cite{ji_02,belitsky_02,boer_03a,collins_03,ji_04a,collins_04,bomhof_06}.
Obviously, the $g_{1T}$-contribution in Eq.~(\ref{e:polnucpt}) is not 
suppressed compared to the $g_T$-term which is already present in the 
collinear approach.
It is worthwhile to note that, in contrast to the target SSA, $p_T$-dependent 
effects are suppressed in the case of the lepton asymmetries.
The result in~(\ref{e:polnucpt}) is IR divergent as well, where the divergent
terms are contained in the function $H$. 
In particular the $p_T$-dependent term in $H$ is also associated with an 
IR divergence.
Provided that this divergence cancels after the inclusion of quark-gluon-quark
correlators, one can perform the $p_T$-integral in~(\ref{e:polnucpt})
and might arrive at a description of the target SSA in terms of ordinary 
integrated correlators.
\\
Concerning the inclusion of quark-gluon-quark correlators we limit ourselves 
here to a short qualitative discussion.
Without detailed algebra one finds that two such correlators can contribute
to the target SSA.
Symbolically these objects can be written as
\begin{equation} \label{e:qgq}
\langle P,S | \, \bar{\psi} \, \gamma^+ A_T^i \, \psi \, | P,S \rangle \,,
\qquad
\langle P,S | \, \bar{\psi} \, \gamma^+ \gamma_5 \, A_T^i \, \psi \, | P,S \rangle \,, 
\end{equation}
with $A_T^i$ denoting the transverse components of the gluon field.
It is possible that upon inclusion of such contributions an IR finite 
result for the target SSA can be obtained.
Here one has to keep in mind that the QCD equations of motion relate 
quark-gluon-quark correlators of the type given in~(\ref{e:qgq}) to 
quark-quark correlators.
Actually, it is quite interesting and promising that a certain linear 
combination of the matrix elements in~(\ref{e:qgq}) is connected to the 
particular combination of $g_T$ and $g_{1T}$ in Eq.~(\ref{e:polnucpt}).
\\

\noindent
To summarize, we have investigated transverse SSAs in inclusive DIS off the 
nucleon which, in general, can be induced by multi-photon exchange between 
the leptonic and the hadronic part of the reaction.
Such SSAs exist for a polarized target as well as a polarized incoming
or outgoing lepton.
We have computed the asymmetry for a transversely polarized lepton beam
and for a transversely polarized nucleon target in the framework of the 
parton model.
So far we have only considered contributions from the quark-quark 
correlator.
In this approach the beam spin asymmetry turns out to be proportional
to the twist-2 unpolarized quark density inside the nucleon.
In contrast, the target SSA is a genuine twist-3 observable which is 
reflected by the appearance of the twist-3 parton density $g_T$.
We have also studied the influence of the transverse motion of the
struck quark.
While such effects are suppressed for the lepton asymmetries, one finds
a leading contribution in the case of the target SSA.
Due to an uncancelled IR divergence our present result for the target 
SSA apparently is incomplete.
However, it is likely that the full leading contribution 
(in the Bjorken limit) can be obtained if quark-gluon-quark correlators 
are also taken into consideration.
\\
The transverse spin asymmetries are of ${\cal O}(\alpha_{em})$ and may 
therefore be small.
Moreover, they are suppressed like $m_{pol}/Q$ with $m_{pol}$ denoting
the mass of the polarized particle. 
At least in the case of the nucleon target SSA this suppression is not 
severe as long as $Q$ is in the region of a few GeV.
Although probably difficult, we think it is definitely worthwhile to 
experimentally explore such transverse SSAs.
Currently, measurements could be performed at CERN, DESY, and at 
Jefferson Lab.
\\

\noindent
{\bf Acknowledgements:}
We would like to thank A. Afanasev and C. Weiss for discussions on the
transverse target asymmetry in DIS and, in particular, for drawing our 
attention to the importance of quark-gluon-quark correlators.
A conversation with X. Jiang was very valuable and helped us to find
earlier work on single spin asymmetries in DIS.
We are also grateful to J.C. Collins and O. Nachtmann for discussions.
This research is part of the EU Integrated Infrastructure Initiative
Hadronphysics Project under contract number RII3-CT-2004-506078.
The work is partially supported by the Verbundforschung ``Hadronen und
Kerne'' of the BMBF.


\end{document}